\documentstyle[epsfig,aps,prb]{revtex}
\def\beq{\begin{equation}}
\def\eeq{\end{equation}}
\def\beqa{\begin{eqnarray}}
\def\eeqa{\end{eqnarray}}
\def\l{\lambda}
\def\e{\eta}
\def\tto{\tau^{-1}}
\def\kk2{\tau^{-2}}
\tightenlines
\begin{document}
\title{
A  square-well model for the structural and 
thermodynamic properties of simple colloidal 
systems}
\author{L.\ Acedo}
\address{
Departamento de F\'{\i}sica, Universidad  de  Extremadura, \\
E-06071 Badajoz, Spain}
\author{A.\ Santos\cite{address}}
\address{
Department of Physics, University of 
Florida, \\
Gainesville, FL 32611}
\date{\today}
\maketitle
\begin{abstract}
A model for the radial distribution function $g(r)$ of a square-well fluid of
variable width 
previously proposed [S. B. Yuste and A. Santos, J. Chem. Phys. {\bf 101}, 2355
(1994)] is revisited and simplified. The model provides an explicit expression
for the Laplace transform of $rg(r)$, the coefficients being given as explicit
functions of the density, the temperature, and the interaction range.
In the limits corresponding to hard spheres and sticky hard spheres the model
reduces to the analytical solutions of the Percus-Yevick equation for those
potentials. The results can be useful to describe in a fully analytical way 
the structural and thermodynamic behavior of colloidal suspensions
modeled as hard-core particles with a short-range attraction. 
Comparison with computer simulation data shows a general good agreement, even
for relatively wide wells. 

\end{abstract}
\draft
 
\section{Introduction}
\label{sec_1}
 In the simplest model of a colloidal dispersion, the interactions among the
(large) solute molecules and the excluded volume effects of the (small) solvent
molecules, which lead to intercolloidal solvation forces, are  incorporated by 
 treating the 
solute particles as hard spheres (HS) of diameter $\sigma$. This allows one to take advantage of 
the  analytical solution of the Percus-Yevick (PY) integral equation for 
the HS potential.\cite{W63,T63} On the other hand, it is well known that the 
solvent particles (e.g., macromolecules) can induce an effective short-range 
attraction between two 
colloidal particles due to an unbalanced osmotic pressure arising from depletion 
of the solvent particles in the region between the two colloidal ones. 
\cite{AO54,V76,CAR00,DBE99}
This explains the widespread use of Baxter's model of sticky hard spheres 
(SHS)\cite{B68} to represent the properties of colloidal suspensions.
\cite{RR89,RR90,RJL90,KRBDVM89,DMVK91,JBH91,MMS91,J95,RAH95,J96,AR97,J98}
This interaction model represents the attractive part of the potential as an 
infinitely deep, infinitely narrow well. In addition to the hard-core diameter 
$\sigma$, the model  incorporates a ``stickyness'' parameter
$\tau_{\text{SHS}}^{-1}$ (essentially the 
deviation of the second virial coefficient from the HS value) that can be 
understood as a measure of the temperature: the smaller the temperature the 
larger the stickyness. The SHS potential has the advantage of lending itself 
to an analytical solution in the PY approximation\cite{B68} and in the Mean Spherical
Approximation (MSA). \cite{BSC87}
On the other hand, the SHS potential presents two limitations. First, the system 
of monodisperse SHS is not thermodynamically stable;\cite{S91,BJS97} however 
this pathology, which is not captured by Baxter's solution to the PY equation, 
can be remedied by including some degree of polydispersity in the system. The 
most important limitation of the SHS model as a representation of a realistic 
short-range interaction lies in the fact that it is unable to distinguish two 
situations with the same ``stickyness'' (i.e. same second virial coefficient) 
and density, but different temperature and/or range.

In order to mimic the particle interaction in colloidal systems in a more 
realistic way, a simple choice is to assume that the particles interact via the
square-well (SW) potential \cite{LKLLW99,DFFGSSTVZ00,ZFDST00}
\begin{equation}
\label{a1}
\varphi(r)=\left\{
\begin{array}{ll}
\infty  ,& r<\sigma \\
-\epsilon  ,& \sigma<r<\lambda \sigma \\
0,&r>\lambda \sigma ,
\end{array}
\right.
\end{equation}
where $\sigma$ is the diameter of the hard core, $\epsilon$ is the well depth,
and $(\lambda-1)\sigma$ is the well width.
The equilibrium properties of a SW fluid depend on the values of three 
dimensionless parameters: the reduced number density $\rho^*=\rho \sigma^3$, the
reduced temperature $T^*=k_BT/\epsilon$ ($k_B$ being the Boltzmann constant), 
and 
the width parameter $\lambda$.
In the limits $\epsilon \rightarrow 0$ (i.e.
$T^*\rightarrow \infty$) and/or $\lambda \rightarrow 1$, the SW fluid becomes 
the HS fluid.
In addition,  the SHS fluid \cite{B68} is obtained 
by taking the 
limits $\epsilon \rightarrow \infty$ (i.e.
$T^*\rightarrow 0$) and $\lambda \rightarrow 1$, while keeping the stickyness parameter
$\tau_{\text{SHS}}^{-1}=12(\lambda-1)(e^{1/T^*}-1)$ 
constant.
If we define in the original SW fluid a generalized stickyness parameter 
$\tau_{\text{SW}}^{-1}=4(\lambda^3-1)\left(e^{1/T^*}-1\right)$ as 
being proportional to the deviation of the second virial 
coefficient from the HS 
value,\cite{NF00} then
 at a given density $\rho^*$ the parameter space  can be
taken as the plane $(\tau_{\text{SW}}^{-1},\lambda)$. The SHS limit explores  the 
line $(\tau_{\text{SW}}^{-1},\lambda=1)$ only, while the HS model corresponds to the 
point $(\tau_{\text{SW}}^{-1},\lambda)=(0,1)$.
This geometrical picture illustrates why the SW interaction model can be 
useful 
to uncover a much richer spectrum of values for the relevant parameters of the 
problem, even if the attraction range is relatively short. 

Despite the mathematical simplicity of the SW potential, no analytical solution 
of the 
conventional integral equations for fluids (YBG, HNC, PY, \ldots) is 
known. \cite{BH76,C96b}
The mean spherical model approximation of Sharma and Sharma\cite{SS77} provides 
an
analytical expression for the structure factor, but it is not consistent with
the hard core exclusion constraint.
Most of the available theoretical information about the SW fluid for variable 
width comes from perturbation 
theory.\cite{SHB70,SHB71,HBS76,HSS80,CS94,TL93,TL94a,TL94b,BG99}
In general, perturbation theory is based on an expansion of the relevant
physical quantities in powers of the inverse temperature. For instance, the
radial distribution function $g(r;\rho^*,T^*)$ of the SW fluid is expressed as
\beq
g(r;\rho^*,T^*)=g_0(r;\rho^*)+{T^*}^{-1}g_1(r;\rho^*)+\cdots,
\label{a2}
\eeq
where $g_0(r;\rho^*)$ is the radial distribution function of the reference HS
fluid and $g_1(r;\rho^*)$ represents a first order correction. Good analytical
approximations for $g_0$ are known, such as Wertheim-Thiele's solution of the PY
equation, \cite{W63,T63} Verlet-Weiss parameterization, \cite{BH76} or the
Generalized Mean Spherical Approximation. \cite{YS91,YHS96} Thus, it is in the choice
of $g_1$ where  different versions of perturbation theory essentially differ. A few
years ago, Tang and Lu (TL)\cite{TL94a,TL94b} proposed an analytical expression
(in Laplace space) for $g_1$, based on the MSA.
Comparison with Monte Carlo (MC) simulation data \cite{HMF76} showed a general good
agreement for the cases considered, but the quality of the agreement worsened as smaller values
of $\lambda$ and/or $T^*$ were taken. 
The expectation that perturbation theory becomes less accurate as the well width
and the temperature
decrease has been already reported elsewhere.\cite{HSS80,CS94} In fact,
perturbation theory tends to overestimate the critical temperature $T_c^*$,
especially for small values of $\lambda-1$.\cite{HSS80,CS94}
All these limitations are significantly apparent in the SHS limit ($\lambda\to
1$, $T^*\to 0$), in which case the expansion (\ref{a2}) becomes meaningless.
By following a completely different approach, Nezbeda \cite{N77} proposed to
approximate $r g(r)$ in the interval $\sigma\leq r\leq \lambda\sigma$ by a
quadratic function of $r$, within the context of the PY theory. The
coefficients of the polynomial were then determined analytically by imposing
the continuity of the cavity function $y(r)\equiv g(r)e^{\varphi(r)/k_BT}$ and its first two
derivatives at $r=\sigma$. Notwithstanding the merits of this theory, its main
limitation is that it is only applicable for very narrow wells, typically
$\lambda-1\lesssim 0.01$.\cite{LKLLW99,N77,LS95} In addition, Nezbeda's
theory fails to predict a thermodynamic critical point, except at $\lambda=1$,
in which case Baxter's solution for SHS is recovered.\cite{N77}

In order to provide a simple theory that, while including the SHS case as a
special limit, could also be applicable to SW fluids of variable width, Yuste
and Santos  proposed a model by assuming an explicit
functional form for the Laplace transform $G(t)$ of $rg(r)$.\cite{YS94} That functional form
was suggested by the exact virial expansion of the radial distribution function
\cite{BH67} and by the property $\lim_{r\rightarrow
\sigma^+}g(r)=\text{finite}$. The parameters were subsequently determined as
functions of $\rho^*$, $T^*$, and $\lambda$ by
imposing the condition $S(0)=\text{finite}$, where $S(q)$ is the structure
function, as well as the continuity of $y(r)$ at $r=\lambda\sigma$. 
The structural properties predicted by
the model showed a good agreement with MC simulation results, not only for narrow
square wells, but even for relatively wide ones ($\lambda\approx 1.5$) up to
densities slightly above the critical density.\cite{YS94} 
On the other hand, the model was not fully analytical because the continuity
condition of $y(r)$ at $r=\lambda\sigma$ gives rise to a transcendent equation that
must be solved numerically. In fact, the {\em exact\/} solution in the case of
one-dimensional SW fluids  involves a similar transcendent
equation.\cite{YS94,SZK53}

The aim of this paper is to propose a simpler version of the model introduced
by Yuste and Santos in Ref.\ \onlinecite{YS94}. While we keep the same
functional structure of $G(t)$ and enforce the conditions
$g(\sigma^+)=\text{finite}$ and $S(0)=\text{finite}$, we replace the
transcendent equation stemming from the continuity of $y(r)$ at
$r=\lambda\sigma$ by a quadratic equation suggested by the SHS limit. The
resulting model has therefore a degree of algebraic simplicity similar to that of the PY
solution for SHS (but now the parameters have a $\lambda$-dependence
beyond the one captured by the stickyness coefficient) and reduces to it in the
appropriate limit.
The structural properties $g(r)$ and $S(q)$ exhibit a fairly good agreement
with MC simulations, \cite{HSS80,HSKGKQ84} similar to that found in
the original, more complicated version of the model.\cite{YS94} Nevertheless,
it is in the calculation of the thermodynamic properties (which were not
addressed in Ref.\ \onlinecite{YS94}) where the present model becomes
especially advantageous. In particular, the isothermal compressibility is
obtained as an explicit function of density, temperature, and well width. This
allows us to get the $\lambda$-dependence of the critical temperature $T_c^*$
and density $\rho_c^*$. Comparison with computer simulation estimates
\cite{VMRJM92} shows that the model predictions for $T_c^*(\lambda)$ are
remarkably good, even for values of $\lambda$ as large as $\lambda=1.75$. On
the other hand, the predicted values of $\rho_c^*(\lambda)$ are typically
30--45\% smaller than the simulation ones, a fact that can be traced back to
the solution of the PY equation for SHS.
We also compare the predictions of the model for the compressibility factor at
$\lambda=1.125$ and $\lambda=1.4$ with simulation data \cite{HSS80,LLA90} and
with the TL perturbation theory.\cite{TL94b} In both cases the model presents a
better agreement than the perturbation theory, except for temperatures larger
than about twice the critical temperature.

        The paper is organized as follows.
The model is introduced and worked out in Sec.\ \ref{sec_2}, with some
technicalities being relegated to the appendices.
Section \ref{sec_3} deals with the comparison with simulation results and
perturbation theory.
The paper ends with a brief discussion in Sec.\ \ref{sec_4}

\section{The model}
\label{sec_2}
 \subsection{Basic requirements}       
        The radial distribution function $g(r)$ of a fluid is directly related to the probability of 
finding two particles separated by a distance $r$.\cite{HM86} It can be 
measured from neutron or x-ray diffraction 
experiments through the static structure factor $S(q)$.
Both quantities are related by Fourier transforms:
\begin{eqnarray}
\label{b1}
S(q)&=&1+\rho \int d{\bf r}\, e^{-i{\bf q}\cdot {\bf r}} [g(r)-1]\nonumber\\
&=&1-2\pi\rho \left.\frac{G(t)-G(-t)}{t}\right|_{t=iq},
\end{eqnarray}
where  $\rho$ is the
number density and 
\begin{equation}
\label{b3}
G(t)=\int_0^\infty dr\, e^{-rt} rg(r)
\end{equation}
is the Laplace transform of $rg(r)$.
The isothermal compressibility of the fluid, $\kappa_T=\rho^{-1}\left(\partial 
\rho/\partial p\right)_{T}$, is directly related to the 
long-wavelength limit of the structure function:
\beq
\chi_T\equiv \rho k_BT\kappa_T=S(0).
\label{b2}
\eeq
Thus, all the physically
relevant  information about the equilibrium state of the system is contained in 
$g(r)$ or, equivalently, in $G(t)$.

Now we particularize to the SW interaction potential, Eq.\ (\ref{a1}), so 
$g(r)=0$ for $r<\sigma$. Henceforth,
we will  take the hard-core diameter $\sigma=1$ as the length unit and the well
depth $\epsilon/k_B=1$ as the temperature unit, so that the asterisks in
$\rho^*$ and $T^*$ will be dropped.
It is convenient to define an 
auxiliary function $F(t)$ through the relation
\begin{eqnarray}
\label{b5}
G(t)&=&t\frac{F(t)e^{-t}}{1+12\eta F(t)e^{-t}}\nonumber\\
 &=&\sum_{n=1}^\infty (-12\eta)^{n-1}t[F(t)]^n e^{-nt},
\end{eqnarray}
where $\eta \equiv \frac{\pi}{6}\rho\sigma^3$ is the packing fraction.
Laplace inversion of Eq.\ (\ref{b5}) provides an useful representation of
 the radial distribution function:
 \begin{equation}
 \label{b6}
 g(r)=r^{-1}\sum_{n=1}^\infty (-12\eta)^{n-1}f_n(r-n)\Theta(r-n),
 \end{equation}
where $f_n(r)$ is the inverse Laplace transform of $t[F(t)]^n$ and $\Theta(r)$
is Heaviside's step function.
Thus, the knowledge of $F(t)$ is fully equivalent to that of $g(r)$
or $S(q)$.
In particular, the value of $g(r)$ at contact, $g(1^+)$, is given by
the asymptotic behavior of $F(t)$ for large $t$:
\begin{equation}
\label{b7}
g(1^+)=f_1(0)=\lim_{t\rightarrow \infty} t^2F(t).
\end{equation}
Since $g(1^+)$ must be finite and different from zero, we get the condition
\begin{equation}
\label{b9}
F(t)\sim t^{-2},\quad t\rightarrow \infty.
\end{equation}
On the other hand, according to Eq.\ (\ref{b1}), the behavior of $G(t)$ for 
small $t$
determines the value of $S(0)$:
\begin{equation}
\label{b8}
G(t)=t^{-2}+\text{const}+\frac{1-S(0)}{24\eta}t+o(t^2).
\end{equation}
Insertion of Eq.\ (\ref{b8}) into the first equality of Eq.\ (\ref{b5}) yields 
the first five terms in the expansion 
of $F(t)$ in
powers of $t$:\cite{YS91,YHS96}
\begin{equation}
\label{b10}
F(t)=-\frac{1}{12\eta}\left(1+t+\frac{1}{2}t^2+\frac{1+2\eta}{12\eta}t^3
+\frac{2+\eta}{24\eta}t^4\right)+{\cal O}(t^5).
\end{equation}
The value of $S(0)$ is related to the coefficients of $t^5$ and $t^6$ by
\beq
S(0)=\frac{24}{5}\eta^3\left[6\left.\frac{d^5F(t)}{dt^5}\right|_{t=0} - 
\left.\frac{d^6F(t)}{dt^6}\right|_{t=0}\right]-1+8\eta+2\eta^2.
\label{b4}
\eeq
So far, all the expressions apply to any density and any hard-core potential.  
As is well known, in the limit of zero density the cavity function $y(r)\equiv 
g(r)e^{\varphi(r)/k_BT}$ is equal to 1.\cite{HM86} In the special case of the 
SW potential this translates into 
\beq
\lim_{\eta\to 0} g(r)=(1+x)\Theta(r-1)-x\Theta(r-\lambda),
\label{b11}
\eeq
where $x\equiv e^{1/T}-1$. Equation (\ref{b11}) implies that
\beq
\lim_{\eta\to 0} F(t)=(1+x)(t^{-2}+t^{-3})-x e^{-(\lambda-1) t}(\lambda 
t^{-2}+t^{-3}).
\label{b12}
\eeq
\subsection{Construction of the model}

Any meaningful approximation of $F(t)$ for the SW potential must comply with 
Eqs.\ (\ref{b9}), (\ref{b10}), and (\ref{b12}). 
Let us decompose $F(t)$ as
\beq
F(t)=R(t)-\overline{R}(t)e^{-(\lambda-1)t}.
\label{c0}
\eeq
The model proposed by Yuste and 
Santos \cite{YS94} consists of assuming the following rational forms for $R(t)$
and $\overline{R}(t)$:
\beq
R(t)=\frac{A_0+A_1 t}{1+
S_1t+S_2 t^2+S_3t^3},\quad \overline{R}(t)=\frac{\overline{A}_0+\overline{A}_1 t}
{1+
S_1t+S_2 t^2+S_3t^3}.
\label{c1}
\eeq
This forms are compatible with (\ref{b12}). In addition,
condition (\ref{b9}) is satisfied by construction.
In fact, the contact value is, according to Eq.\ (\ref{b7}),
\beq
g(1^+)=\frac{A_1}{S_3}.
\label{c4}
\eeq
In order to ease the proof that  the HS and SHS cases are included in
Eq.\ (\ref{c1}), it is convenient to introduce the new parameters $A\equiv
-12\eta \overline{A}_0$, $L_1\equiv
-12\eta\left[\overline{A}_0(\lambda-1)+A_1-\overline{A}_1\right]$, 
and $L_2\equiv -12\eta
\overline{A}_1(\lambda-1)$. With these changes, Eqs.\ (\ref{c0}) and (\ref{c1}) can be recast
into
\begin{equation}
\label{c3}
F(t)=-\frac{1}{12\eta}\frac{1+A+[L_1+L_2(\l-1)^{-1}-A(\l-1)] t-[{A}
+L_2(\lambda-1)^{-1}t]e^{-(\lambda-1)t}}{1+
S_1t+S_2 t^2+S_3t^3},
\end{equation}
where we have already taken into account the property $F(0)=-1/12\eta$,
according to Eq.\ (\ref{b10}).
The model (\ref{c3}) contains six parameters to be determined. 
The exact expansion (\ref{b10}) imposes four constraints among them.
Thus, we can express four of the parameters in terms of, for instance, 
${A}$ and $L_2$. The result is
\begin{equation}
\label{c5}
L_1=\frac{1}{1+2\eta}\left[1+\frac{1}{2}\eta+2\eta(1+\lambda+\l^2) 
L_2-\frac{1}{2}\eta(3+2\l+\l^2)
{A'}\right],
\end{equation}
\begin{equation}
\label{c6}
S_1=\frac{\eta}{1+2\eta}\left[-\frac{3}{2}+2(1+\lambda+\l^2) L_2-
\frac{1}{2}(3+2\l+\l^2)
{A'}\right],
\end{equation}
\beq
\label{c7}
S_2=\frac{1}{2(1+2\eta)}\left\{-1+\eta+2[1-2\eta\l(1+\lambda)]L_2
-[1-\eta(1+\l)^2]{A'}\right\},
\eeq
\beq
\label{c7bis}
S_3=\frac{1}{1+2\eta}\left\{-\frac{(1-\eta)^2}{12\eta}-
\left[\frac{1}{2}(\lambda+1)-\eta\lambda^2\right]L_2
+\frac{1}{12}[4+2\l-\eta(3\l^2+2\l+1)]{A'}\right\},
\eeq
where $A'\equiv A(\l-1)^2$.
These four parameters are linear functions of $A$ and $L_2$. Taking into account 
Eq.\ (\ref{b4}), the value of $S(0)$ can be expressed as a quadratic function of 
$A$ and $L_2$. Its explicit expression is given in Appendix \ref{appA}.
  
Two additional constraints are still needed to determine $A$ and $L_2$.
First note that in the zero-density limit we have $L_1\to 1$, $S_1\to 0$, 
$S_2\to \text{finite}$, $S_3\to -(12\e)^{-1}$. Thus, Eq.\ (\ref{c3}) is 
consistent with Eq.\ (\ref{b12}) provided that
\beq
\lim_{\eta\to 0} A=x,\quad \lim_{\eta\to 0} L_2=x\lambda(\lambda-1).
\label{c9}
\eeq
In the original formulation of the model,\cite{YS94} the parameter $A$ was 
assumed to be independent of density, so that $A=x$. As for $L_2$, it was 
determined by imposing the (exact) continuity condition of the
function $y(r)$ at $r=\lambda$,\cite{BH76,A00} which implies
\begin{equation}
g(\lambda^-)=(1+x)g(\lambda^+).
\label{c8}
\end{equation}
The implementation of this condition in the model (\ref{c3}) leads to a 
transcendent equation that must be solved numerically.\cite{YS94} In this paper 
we will be concerned with a simpler version of the model in which the strong 
condition (\ref{c8}) is replaced by a weaker one. To that end, let us 
introduce a parameter $\tau$ as
\beq
\tau\equiv \frac{1}{12}\left(\frac{L_1}{L_2}-\frac{S_2}{S_3}\right).
\label{c10}
\eeq
{}From (\ref{c9}) and (\ref{c10}) it follows that
\beq
\lim_{\eta\to 0} \tau=[12 x\lambda(\lambda-1)]^{-1}.
\label{c12}
\eeq
The definition (\ref{c10}) is suggested by the fact that, as proved in Appendix B of Ref.\ 
\onlinecite{YS94}, 
Eq.\ (\ref{c8}) is equivalent to $\tau=\tau_{\text{SHS}}$ in the SHS limit, 
namely $x\to\infty$, $\lambda\to 1$, 
$x(\lambda-1)\equiv(12\tau_{\text{SHS}})^{-1}=\text{finite}$. 
Therefore, we may expect that a simple prescription for $\tau$ (such that
$\tau^{-1}\to 12x(\lambda-1)$ in the SHS limit) can be a good 
substitute for the transcendent equation arising from (\ref{c8}), at least for 
 wells relatively narrow.
  
Since $L_1$, $S_2$, and $S_3$ are linear functions of $L_2$, Eq.\ (\ref{c10}) 
shows that $\tau$ is the ratio of two quadratic functions of $L_2$. This 
relation is easily inverted to get
\beq
L_2=\frac{-1+\alpha_1 \eta+\alpha_2 \eta^2+\alpha_3 
\eta^3+(1+2\e)\left[1+\beta_1 \eta+\beta_2 \eta^2+\beta_3 \eta^3+\beta_4 
\eta^4\right]^{1/2}}{12\e(\gamma_0+\gamma_1 \eta+\gamma_2 \eta^2)},
\label{c11}
\eeq
where the expressions of the coefficients $\alpha_i$, $\beta_i$, and $\gamma_i$ 
as functions of $\lambda$, $A$, and $\tau$ are given in Appendix \ref{appA}.

So far, we are free to fix $\tau$ as a function of $\eta$, $x$, and $\lambda$ by
following any criterion we wish. In particular, we can enforce Eq.\ (\ref{c8}),
as done in Ref.\ \onlinecite{YS94}. Analogously, the parameter
$A(\eta,x,\lambda)$ (which was
taken $A=x$ in Ref.\ \onlinecite{YS94}) can be freely chosen.
In the simplified version of the model we consider here, we assume that both $A$ 
and $\tau$ are {\em independent\/} of density, so they take values needed to 
satisfy the requirement (\ref{b12}). Those values are simply
\beq
A=x,\quad \tau=[12 x\lambda(\lambda-1)]^{-1}.
\label{c13}
\eeq
Note that this $\tau$ is only slightly different from the parameter
$\tau_{\text{SW}}$ introduced below Eq.\ (\ref{a1}), both of them becoming
identical to  $\tau_{\text{SHS}}$ in the limit $\lambda\to 1$.	
The choice (\ref{c13}) closes the construction of the model. The pair distribution function in 
Laplace space is given by Eqs.\ (\ref{b5}) and (\ref{c3}), where the expressions 
for the coefficients are (\ref{c5})--(\ref{c7bis}), (\ref{c11}), and 
(\ref{c13}). The model provides the quantity $G(t)$ as an explicit function of 
the Laplace variable $t$, the packing fraction $\eta$, the well width $\l$, and 
the temperature parameter $x\equiv e^{1/T}-1$. 
Since the poles of $F(t)$ are the roots of a cubic equation, the inverse Laplace 
transforms of $t[F(t)]^n$ are analytically derived and then the radial 
distribution function is readily obtained from the representation (\ref{b6}). 
{}From Eq.\ (\ref{c13}) it follows that the relationship between the temperature 
and the parameter $\tau$ is
\beq
T=1/\ln[1+\tto/12\l(\l-1)].  
\label{d6}
\eeq

The predictions of our model to first order in the packing fraction are compared 
with the exact results in Appendix \ref{appAB}. 
As an illustrative example, Fig.\ \ref{fig0} shows the quantity
\beq
\Delta(r)=\lim_{\eta\to0}\frac{1}{\eta}\frac{g(r)-g_{\text{exact}}(r)}{g_{\text{exact}}(r)}
\label{d6.1}
\eeq
for $\lambda=1.1$ and $\lambda=1.125$ and for the temperatures $T=0.5$, 0.67,
and 1.
The function $\Delta(r)$ is different from zero in the interval $1<r<\lambda$ only,
where our model slightly overestimates the value of $g(r)$. Note that the
relative deviation of our $g(r)$ from the exact radial distribution function is
$\Delta(r)\eta$ for small densities. Of course, the linear growth of this
deviation with increasing density is valid in this low-density regime only, as
comparison with simulation values at finite densities shows [cf.\ Figs.\
\ref{fig3} and \ref{fig4}].
\begin{figure}
\begin{center}\parbox{\textwidth}{\epsfxsize=0.7\hsize\epsfbox{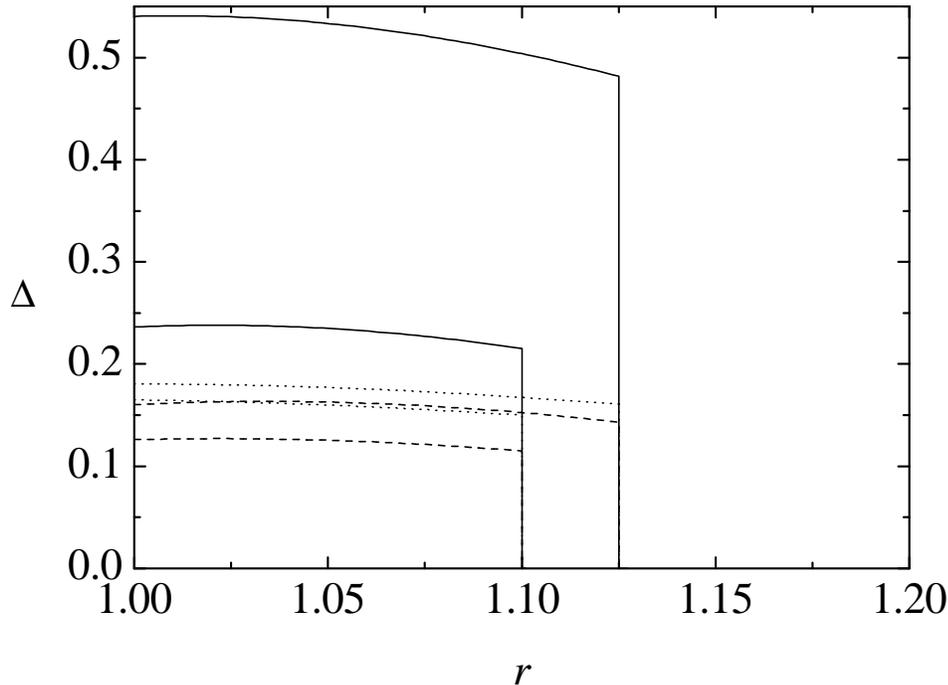}}
\caption{Function $\Delta(r)$ defined by Eq.\ (\protect\ref{d6.1}) for 
$\lambda=1.1$ and $\lambda=1.125$  and for the temperatures $T=0.5$ (solid
lines), 0.67 (dashed lines),
and 1 (dotted lines).
\label{fig0}}
\end{center}
\end{figure}

Appendix \ref{appB} shows that 
the model reduces to Wertheim-Thiele's and 
Baxter's analytical solutions of the PY equation in the HS and SHS limits, 
respectively.
If we consider the parameter space $\l$--$\tto$, then the HS potential 
corresponds to the line $\tto=0$ (in which case the physical properties are 
independent of $\l$), while the SHS potential corresponds to the line $\l=1$ 
(the physical properties being $\tau$-dependent). What our model does is to 
extend the above picture to the entire plane $\tto\geq 0$, $\l\geq 1$, without 
compromising the mathematical simplicity present in the analytical solutions of 
the PY integral equation for HS and SHS.

\section{Comparison with simulation and other theories}
\label{sec_3}
\subsection{Structural properties}
\label{sec3.1}
The structural properties obtained from the original version of the model [i.e. 
with $L_2$ determined by solving the transcendent equation stemming from Eq.\ 
(\ref{c8})] were profusely compared with simulation data 
\cite{HSS80,HMF76,HSKGKQ84} in Ref.\ \onlinecite{YS94}. We have checked that the 
simplified version of the model presented here [cf.\ (\ref{c11}), (\ref{c13})] 
gives results  very close to those of the original model, especially 
for narrow wells. Consequently, we will present only a brief comparison with 
simulations in this Subsection.

\begin{figure}
\begin{center}\parbox{\textwidth}{\epsfxsize=0.7\hsize\epsfbox{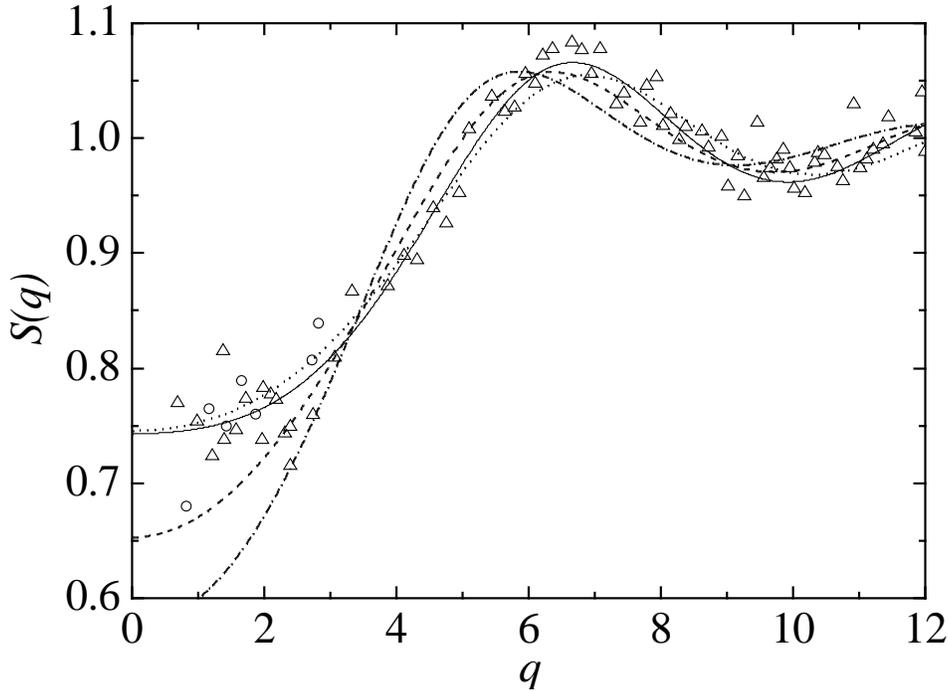}}
\caption{Structure factor, $S(q)$, corresponding to a SW fluid with $\lambda=
1.1$, $\eta=0.07$, and ${T}^{-1}=0.92$.
The circles and triangles are MC data taken from Fig.\ 3 of Ref.\ 
\protect\onlinecite{HSKGKQ84}.
The  lines are the result predicted by the present model (---), the PY equation 
for SHS ($\cdots$), the TL perturbation theory (- - -), and the PY equation 
for HS (-$\cdot$-$\cdot$-).
\label{fig1}}
\end{center}
\end{figure}
\begin{figure}
\begin{center}\parbox{\textwidth}{\epsfxsize=0.7\hsize\epsfbox{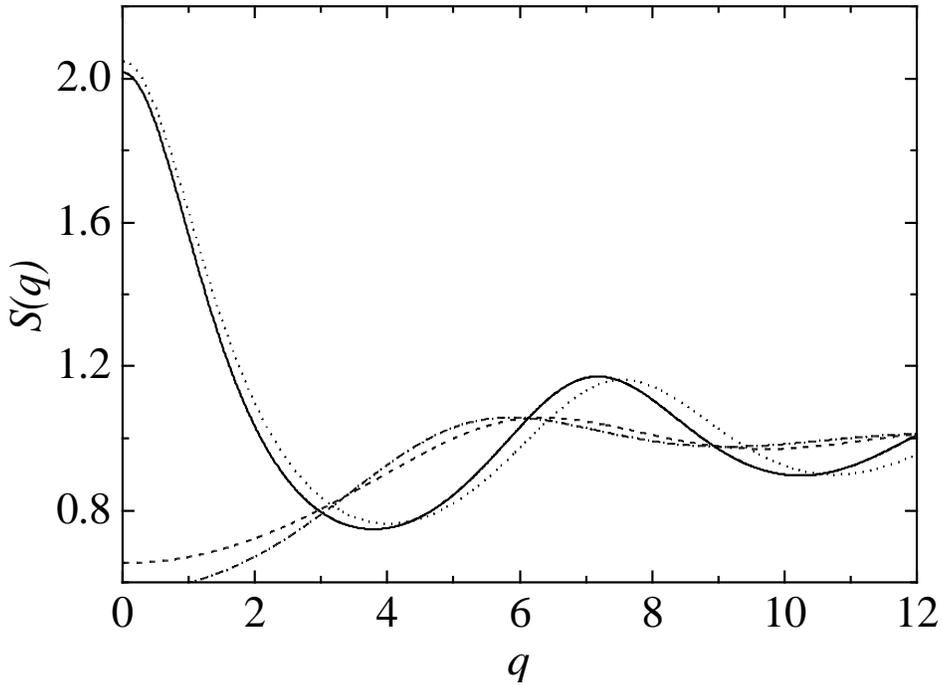}}
\caption{Structure factor, $S(q)$, corresponding to a SW fluid with $\lambda=
1.1$, $\eta=0.07$, and ${T}=0.5$.
The  lines are the result predicted by the present model (---), the PY equation 
for SHS ($\cdots$), the TL perturbation theory (- - -), and the PY equation 
for HS (-$\cdot$-$\cdot$-).
\label{fig2}}
\end{center}
\end{figure}

The structure of a fluid is usually determined by neutron or 
x-ray scattering experiments, which measure the structure factor $S(q)$. This 
quantity is directly related to the Laplace transform $G(t)$ by Eq.\ (\ref{b1}) 
and so can be obtained explicitly in our model. 
In 1984, Huang {\em et al.} \cite{HSKGKQ84} performed MC simulations of SW 
fluids
with $\lambda=1.1$ to reproduce the main features of the
structure factor of micellar suspensions.
Figure \ref{fig1} shows $S(q)$ obtained from simulation \cite{HSKGKQ84} for 
$\lambda
=1.1$, ${T}^{-1}=0.92$, and $\eta=0.07$, as compared with our model, the PY 
solution for 
SHS (with the conventional choice of $\tau^{-1}_{\text{SW}}$ as stickyness
parameter), the Tang-Lu (TL)
perturbation theory,\cite{TL94a,TL94b} and the PY solution for hard spheres.
The deviations of the simulation data from the PY-HS curve are essentially a 
measure of 
effects associated with the non-zero values of the square well width and 
of the inverse temperature. These are qualitatively described by the TL 
perturbation theory, except for small wavenumbers. This region is well 
represented 
by the PY-SHS curve and by our model, but the latter is better near the first 
maximum.
The value of the critical temperature for $\lambda=1.1$ can be estimated to be 
$T_c\approx 0.5$ [cf.\ Table \ref{table1}]. Consequently, the case considered in 
Fig.\ 
\ref{fig1} corresponds to a rather hot gas.  That is why the simulation data are 
not too far from the HS values (except in the region $q\lesssim 3$). In order to 
highlight the differences that can 
be expected at a smaller temperature, the case $T=0.5$ is considered in 
Fig.\ \ref{fig2}. Now, the structure factor predicted by our model and by the 
PY-SHS solution is very different from that predicted by the TL perturbation 
theory, the latter being very close to the PY-HS solution. This is not 
surprising, since in any perturbation theory the quantities are expanded in 
powers of ${T}^{-1}$ and obviously the value ${T}^{-1}=2$ is beyond its 
range of applicability. On the other hand, the curves corresponding to the 
PY-SHS solution and our model are relatively close (except for a slight phase
shift), as expected from the fact 
that the well width is rather small. Moreover, 
since the stickyness parameter has been chosen as to reproduce the correct
second virial coefficient, the PY-SHS solution does a general good job at this
very low  density.

\begin{figure}
\begin{center}\parbox{\textwidth}{\epsfxsize=0.7\hsize\epsfbox{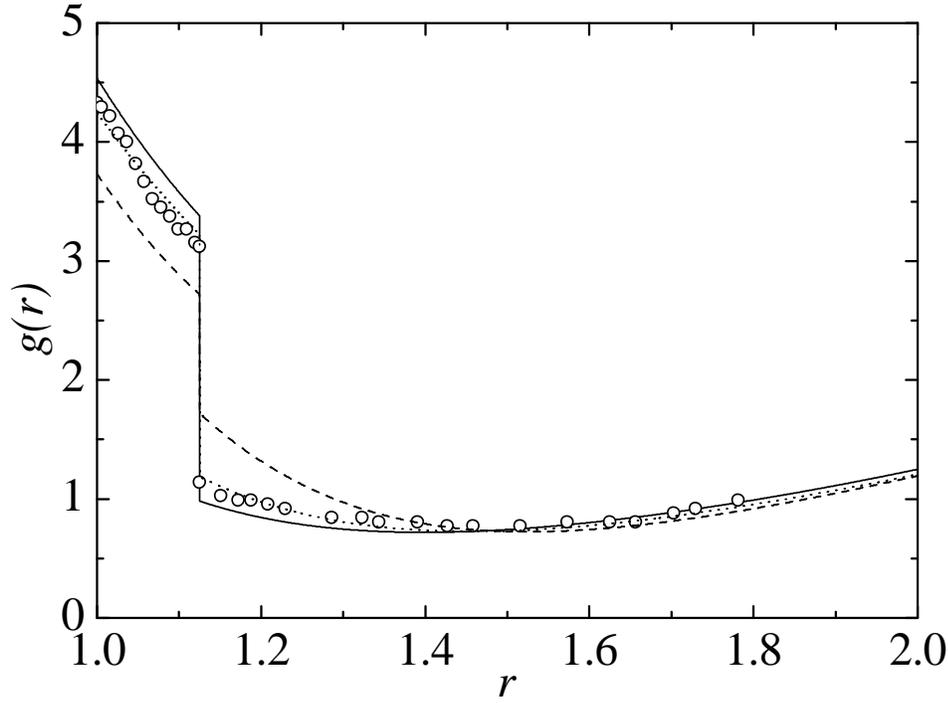}}
\caption{Radial distribution function, $g(r)$, corresponding to a SW fluid with 
$\lambda=1.125$, $\rho=0.8$, and $T=1$. The circles represent MC 
data,\protect\cite{HSS80} 
the dotted line is the result obtained from the
model of Ref.\ \protect\onlinecite{YS94}, the solid line 
is the result given by the present model, and the dashed line is the prediction 
from the 
TL perturbation theory.\label{fig3}}
\end{center}
\end{figure}

\begin{figure}
\begin{center}\parbox{\textwidth}{\epsfxsize=0.7\hsize\epsfbox{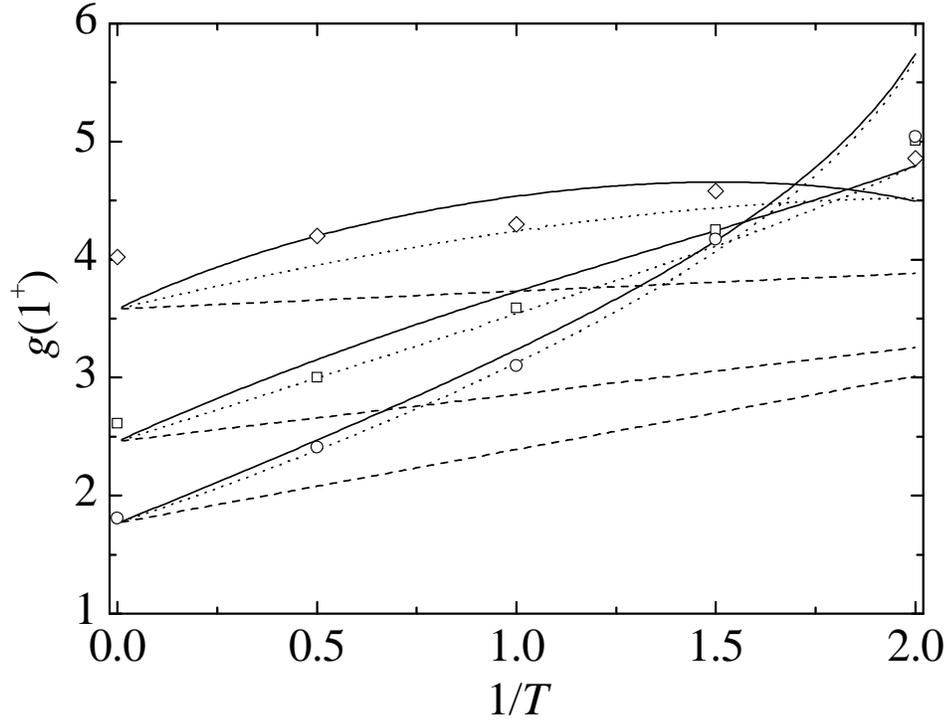}}
\caption{Plot of $g(1^+)$ as a function of $1/T$ for $\lambda=1.125$ and 
three densities: $\rho=0.4$ ($\circ$), $\rho=0.6$ ($\Box$), and $\rho=0.8$ 
($\Diamond$).
The symbols represent MC data taken from Ref.\ \protect\onlinecite{HSS80} 
($1/T\neq 0$)
and from Ref.\ \protect\onlinecite{BH71} ($1/T=0$).
The  the dotted lines are the results obtained from the
model of Ref.\ \protect\onlinecite{YS94}, the solid lines are the results given by the
 present model, and the dashed lines 
are the 
predictions from the TL perturbation theory.
\label{fig4}}
\end{center}
\end{figure}

Now we consider the radial distribution function itself. As a representative 
example of a  width not extremely small, we take the case $\lambda=1.125$, for which 
MC simulations are available.\cite{HSS80} Figure \ref{fig3} shows 
$g(r)$ for $\lambda=1.125$ at the thermodynamic state $T=1$, $\rho=0.8$. 
As seen in the figure, the original model of Ref.\ \onlinecite{YS94} exhibits a remarkable
agreement with the simulation data. We also
observe that our present model captures reasonably well the behavior of $g(r)$ at this
high density, while 
the TL approximation predicts a too small contact value $g(1^+)$.  The 
contact values are plotted in Fig.\ \ref{fig4} as a function of $1/T$ for 
$\lambda=1.125$ and $\rho=0.4$, 0.6, and 0.8. At $1/T=0$ the system 
corresponds to HS and then our model and the TL theory reduce to 
Wertheim-Thiele's solution of the PY equation, which tends to underestimate 
$g(1^+)$ at high densities. As the inverse temperature increases, the TL theory 
predicts a linear increase of $g(1^+)$ [cf.\ Eq.\ (\ref{a2})] that clearly deviates from the MC 
data. On the other hand, the temperature dependence of $g(1^+)$ is well 
described by  both the original and the simplified versions of the model,
 except at $1/T=2$.

\subsection{Thermodynamic properties}
\label{sec3.2}
The compressibility equation of state is obtained from Eqs.\ (\ref{b2})  and 
(\ref{A1}). This route is preferable to the virial route because the latter is 
known to yield an unphysical critical behavior  in the SHS limit.\cite{FF81} 
{}From Eq.\ 
(\ref{b2}), we have
\beq
Z(\eta,T)\equiv \frac{p}{\rho k_BT}=\eta^{-1}\int_0^\eta d\eta'\, 
\chi_T^{-1}(\eta',T),
\label{d1}
\eeq
where the density dependence of $\chi_T$ is given in our model by Eqs.\ 
(\ref{A1}) and (\ref{c11}). Although this dependence is known explicitly, it 
does not allow us to perform the integration in Eq.\ (\ref{d1}) analytically, so 
that the compressibility factor $Z$ is obtained by numerical integration.

\begin{figure}
\begin{center}\parbox{\textwidth}{\epsfxsize=0.7\hsize\epsfbox{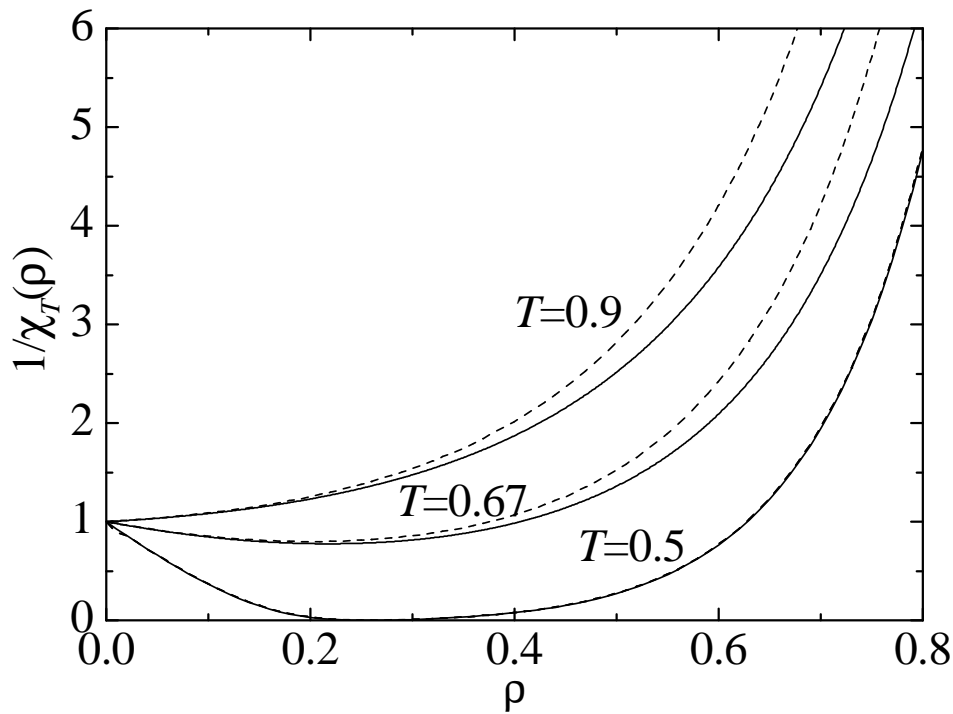}}
\caption{Density dependence of the inverse isothermal susceptibility
 for $\lambda=1.125$ 
and three temperatures. The dashed lines are the results obtained from the
model of Ref.\ \protect\onlinecite{YS94} and  the solid lines are the results from the 
present 
model\label{fig9}}
\end{center}
\end{figure}

 Before delving into the thermodynamic predictions of our model, let us
compare it with the original one \cite{YS94} in the case of a moderately
narrow well, namely $\lambda=1.125$. Figure \ref{fig9} shows the density
dependence of the inverse susceptibility for $T=0.5$, 0.67, and 0.9. As
the temperature  increases, our simplified model is seen to
overestimate the compressibility of the fluid, especially for large densities.
On the other hand, at $T=0.5$ (which is practically the critical temperature, cf.\ Table
\ref{table1}), both versions of the model yield undistinguishable results. 
This is quite encouraging, since it is in the domain of low temperatures (or,
equivalently, high stickyness) where our model is expected to 
 correct the deficiencies of
perturbation theories and become useful.

One of the advantages of our simplified model is that it allows us to
derive explicit expressions for the coordinates of the critical point.
The spinodal line is the locus of points where the isothermal compressibility 
diverges. Equation (\ref{A1}) clearly shows that $\chi_T=S(0)\to \infty$ if and 
only if $L_2\to\infty$. Thus,  according to Eq.\ (\ref{c11}), the spinodal line 
is given by a solution to the quadratic equation 
$\gamma_0+\gamma_1\e+\gamma_2\e^2=0$.
This equation has two real solutions, $\eta_{\pm}(\tau)$, only if $\tau$ is a 
smaller than a critical value $\tau_c$ given by
\beq
\tto_c=12\frac{3+\l+2\sqrt{2\l}}{9-2\l+\l^2}.
\label{d3}
\eeq
The critical temperature $T_c$ is obtained by setting $\tau=\tau_c$ in  Eq.\ 
(\ref{d6}). 
If $\tau<\tau_c$, the smallest root, $\eta_-(\tau)$, does not define the vapor 
branch of the spinodal line because it is also a root of the numerator of Eq.\ 
(\ref{c11}) and so $L_2$ remains finite at $\eta=\eta_-(\tau)$. Therefore, only 
the liquid branch of the spinodal line, $\eta_+(\tau)$, exists. It is given by
\beq
\eta_+(T)=\frac{ (1+\l+\l^2)\sqrt{1 
-\frac{1}{6}(3+\l)\tto+\frac{1}{144}(9-2\l+\l^2)\kk2}+\frac{1}{12}(3+\l^3)\tto
-1-\l+\l^2}{\l\left[\frac{1}{3}(2
+3\l+\l^2+\l^3)\tto-4\l\right]}.
\label{d2}
\eeq
The critical density is $\eta_c=\eta_+(\tau_c)$, i.e.
\beq
\eta_c=\frac{\frac{1}{12}(3+\l^3)\tto
-1-\l+\l^2}{\l\left[\frac{1}{3}(2
+3\l+\l^2+\l^3)\tto-4\l\right]}.
\label{d4}
\eeq
As $\l$ increases,  $T_c$ increases, while $\eta_c$ decreases.
Equations (\ref{d3})--(\ref{d4}) generalize to $\lambda>1$ the results 
predicted 
by the PY solution for SHS,\cite{B68}
\beq
\eta_{+}=\frac{\sqrt{9-6\tto+\kk2/2}+\tto-3}{7\tto-12},\quad
\tto_c=3(2+\sqrt{2})\simeq 10.24, \quad
\eta_c=\frac{3}{2}\sqrt{2}-2\simeq 0.1213.
\label{d5}
\eeq

\begin{figure}
\begin{center}\parbox{\textwidth}{\epsfxsize=0.7\hsize\epsfbox{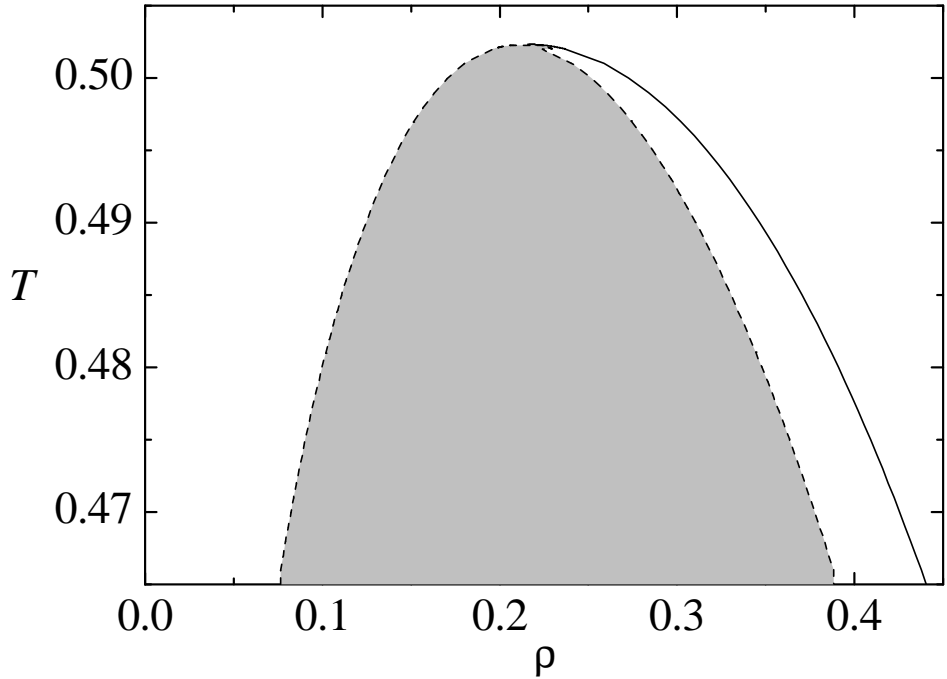}}
\caption{Liquid branch of the spinodal line (solid line) and existence line 
(dashed line) for $\lambda=1.125$, according to the present model. The model 
does not give real values for the physical quantities inside the grey region 
enclosed by the existence line. 
\label{fig5}}
\end{center}
\end{figure}

The lack of a vapor branch of the spinodal line is a feature that our model 
inherits from the PY-SHS solution.\cite{FF81} Our model also has in common with the SHS 
limit, as well as with the solution of the PY equation for finite $\lambda$, 
\cite{T73} the existence of regions in the temperature--density plane, 
inside which the physical quantities cease to take real values. According to 
Eq.\ (\ref{c11}), this happens when 
$1+\beta_1\eta+\beta_2\eta^2+\beta_3\eta^3+\beta_4\eta^4<0$. 
Let us call $\eta_i(T)$ ($i=1,\ldots,4$) the four roots of the quartic equation  
$1+\beta_1\eta+\beta_2\eta^2+\beta_3\eta^3+\beta_4\eta^4=0$, with the convention 
that, whenever they are real, $\eta_1\leq\eta_2\leq\eta_3\leq\eta_4$. It turns
out that the roots 
$\eta_1(T)$ and $\eta_2(T)$ are real only
 if $\tau$ is smaller than a certain threshold value $\tau_{\text{th}}$ 
(or, equivalently, if $T<T_{\text{th}}$); they define a dome-shaped curve with 
an apex at a density 
$\eta_{\text{th}}=\eta_1(T_{\text{th}})=\eta_2(T_{\text{th}})$. 
Analogously, the other two roots, $\eta_3(T)$ and $\eta_4(T)$, are real only
 if $\tau<\tau_{\text{th}}'$ 
($T<T_{\text{th}}'$) and define another dome-shaped curve with an apex at  
$\eta_{\text{th}}'=\eta_3(T_{\text{th}}')=\eta_4(T_{\text{th}}')$. 
Consequently, the parameter $L_2$ becomes complex inside the intervals 
$\eta_1(T)<\eta<\eta_2(T)$ (for $T<T_{\text{th}}$) and 
$\eta_3(T)<\eta<\eta_4(T)$ (for $T<T_{\text{th}}'$). However, the existence of 
the second region is a mathematical artifact since it affects unphysically high 
densities. For instance, for $\lambda=1.125$ we have $T_{\text{th}}'=0.535147$ 
and
$\eta_{\text{th}}'=0.78692$, this  density being larger than the one 
corresponding to the close packing value 
$\eta_{\text{cp}}=\sqrt{2}\pi/6\simeq 0.740$. In fact, in the SHS limit ($\lambda\to 
1$) the 
second region collapses into the line $\eta=1$ and disappears.
On the other hand, the region $\eta_1(T)<\eta<\eta_2(T)$ is inside the curve 
$\eta=\eta_\pm(T)$ and persists in the SHS limit. In the latter case, the 
threshold values  coincide with the critical ones, i.e. 
$\tau_{\text{th}}=\tau_c$, $\eta_{\text{th}}=\eta_c$. For finite $\lambda$, 
$T_{\text{th}}$ is smaller but practically indistinguishable from $T_c$ and 
$\eta_{\text{th}}$ is slightly smaller than $\eta_c$. In the illustrative case of 
$\lambda=1.125$, we have $T_c=0.502324$, $\eta_c=0.11299$, 
$T_{\text{th}}=0.502269$, and $\eta_{\text{th}}=0.10980$.
The existence line, $\rho=\rho_{1,2}(T)$, and the liquid branch of the spinodal 
line, 
$\rho=\rho_+(T)$, are shown in Fig.\ \ref{fig5} for $\lambda=1.125$.

Recently, it has been suggested \cite{VL00} that, while the critical point is in 
general very sensitive to the range of the interaction, the second virial 
coefficient has a fairly constant value at the critical temperature, 
$B_2(T_c)\approx -\pi$. In the case of the SW interaction,   computer 
simulations \cite{VMRJM92,EH99} show that the value of the second virial 
coefficient 
$B_2(T)=-2\pi[x(\l^3-1)-1]/3$ at $T=T_c$ is much less sensitive to the width 
$\lambda$ than the critical temperature.\cite{NF00,VL00}  Moreover, the PY-SHS solution predicts 
$B_2(T_c)=-(2+3\sqrt{2})\pi/6\simeq -1.04 \pi$. Thus, by assuming that 
$B_2(T_c)$ is independent of $\l$ and is equal to its value in the SHS limit, 
the criterion of Ref.\ \onlinecite{VL00} allows one to estimate the critical 
temperature as
\beq
T_c=1/\ln[1+3(2+\sqrt{2})/4(\l^3-1)].  
\label{d7}
\eeq

Since the model presented in this paper is constructed as a simple 
generalization of Baxter's solution of the PY equation for SHS, we expect its 
predictions for the critical values of the temperature ($T_c$), the density 
($\rho_c$), and the compressibility factor ($Z_c$) to be more reliable for small 
values of $\lambda-1$ than for wide wells.  However, to the best of our 
knowledge, the simulation estimates for those quantities are only available for 
$\lambda\geq 1.25$, \cite{VMRJM92} and so we cannot make a comparison for
smaller widths. Table \ref{table1} shows the values of $T_c$, $\rho_c$, and $Z_c$ 
for several values of $\lambda$, as estimated from computer 
simulations,\cite{VMRJM92} and as predicted by the model, by Eq.\ (\ref{d7}), by 
perturbation theory,\cite{HSS80,CS94} and by the PY integral equation.\cite{T73} 
\begin{table}[tbp]
\caption{Critical constants of the square-well fluid for several values of the width parameter $\lambda$. 
}
\label{table1}
\begin{tabular}{lllll}
$\lambda$ & $T_c$& $\rho_c$& $Z_c$&Source\\ 
\tableline
1&0\tablenote{$\tau_c=0.0976$}&0.232&0.379&PY equation\cite{B68}\\
\tableline
1.1&0.455&0.219&0.372&This work\\
    &0.461& & & Eq.\ (\protect\ref{d7})\cite{VL00}\\
\tableline
1.125&0.502&0.216&0.370&This work\\
       &0.512& & & Eq.\ (\protect\ref{d7})\cite{VL00}\\
      &0.594&0.46&0.42&Perturbation theory\cite{HSS80}\\
      &0.587&0.71&0.74&Perturbation theory\cite{CS94}\\
\tableline
1.25&0.764&0.370&0.29&Computer simulation\cite{VMRJM92}\\
 &0.729&0.203&0.360&This work\\
 &0.766& & & Eq.\ (\protect\ref{d7})\cite{VL00}\\
 &0.913&0.34&0.43&Perturbation theory\cite{HSS80}\\
 &0.850&0.48&0.47&Perturbation theory\cite{CS94}\\
\tableline
1.375&0.974&0.355&0.30&Computer simulation\cite{VMRJM92}\\
 &0.960&0.193&0.349&This work\\
 &1.046& & & Eq.\ (\protect\ref{d7})\cite{VL00}\\
  &1.11&0.34&0.39&Perturbation theory\cite{HSS80}\\
  &1.08&0.36&0.40&Perturbation theory\cite{CS94}\\
\tableline
1.5&1.219&0.299&0.30&Computer simulation\cite{VMRJM92}\\
 &1.209&0.184&0.339&This work\\
 &1.367& & & Eq.\ (\protect\ref{d7})\cite{VL00}\\
 &1.205&0.200&0.37&PY equation\cite{T73}\\ 
 &1.35&0.31&0.36&Perturbation theory\cite{HSS80}\\
 &1.33&0.29&0.37&Perturbation theory\cite{CS94}\\
\tableline
1.625&1.479&0.177&0.330&This work\\
 &1.738& & & Eq.\ (\protect\ref{d7})\cite{VL00}\\
 &1.70&0.27&0.38&Perturbation theory\cite{HSS80}\\
 &1.61&0.26&0.36&Perturbation theory\cite{CS94}\\
\tableline
1.75&1.811&0.284&0.35&Computer simulation\cite{VMRJM92}\\
 &1.777&0.170&0.322&This work\\
 &2.164& & & Eq.\ (\protect\ref{d7})\cite{VL00}\\
 &2.04&0.25&0.38&Perturbation theory\cite{HSS80}\\
 &1.93&0.24&0.36&Perturbation theory\cite{CS94}\\
\tableline
1.85& 2.036&0.165&0.317&This work\\        
 &2.550& & & Eq.\ (\protect\ref{d7})\cite{VL00}\\
 &2.33&0.25&0.37&Perturbation theory\cite{HSS80}\\
 &2.23&0.23&0.35&Perturbation theory\cite{CS94}\\
\tableline
2&2.764&0.225&0.32&Computer simulation\cite{VMRJM92}\\
 & 2.466&0.159&0.310&This work\\  
 &3.208& & & Eq.\ (\protect\ref{d7})\cite{VL00}\\
 &2.88&0.24&0.37&Perturbation theory\cite{HSS80}\\
 &2.79&0.23&0.35&Perturbation theory\cite{CS94}
\end{tabular}
\end{table}

Comparison with MC simulation data shows that the critical temperature predicted 
by our model, Eq.\ (\ref{d3}), tends to be smaller than the correct value, while 
Vliegenthart and Lekkerkerker's criterion, \cite{VL00} as well as second-order 
perturbation theory\cite{HSS80,CS94} tend to overestimate it.
What is indeed remarkable is the fact that Eq.\ (\ref{d3}) provides the best 
agreement with computer simulations for $\lambda\leq 1.75$ (except at 
$\lambda=1.25$, in which case Eq.\ (\ref{d7}) is better). In the case of the 
critical value of the compressibility factor, the best agreement corresponds to 
our model, except at $\lambda=1.75$, where Chang and Sandler's perturbation 
theory gives a better result.
On the other hand, the critical densities predicted by our model are around 
30--45\% smaller than the simulation values, a discrepancy that can be traced 
back to the solution of the PY equation for SHS. Since the computer simulations 
and all the theories share the property that $\rho_c$ is a monotonically 
decreasing function of $\lambda$, we can conclude that the correct value of 
$\rho_c$ in the limit $\lambda\to 1$ is certainly larger than the simulation
value $\rho_c=0.370$ at $\lambda=1.25$, while  
Baxter's solution yields $\rho_c=0.232$. Thus, we can expect this failure to 
reproduce accurately the critical density is also present in the PY 
approximation for finite $\lambda$. This is confirmed by the results of a 
numerical solution of the PY equation for $\lambda=1.5$,\cite{T73} which gives a 
value of $\rho_c$ rather close to the one obtained here.

\begin{figure}
\begin{center}\parbox{\textwidth}{\epsfxsize=0.7\hsize\epsfbox{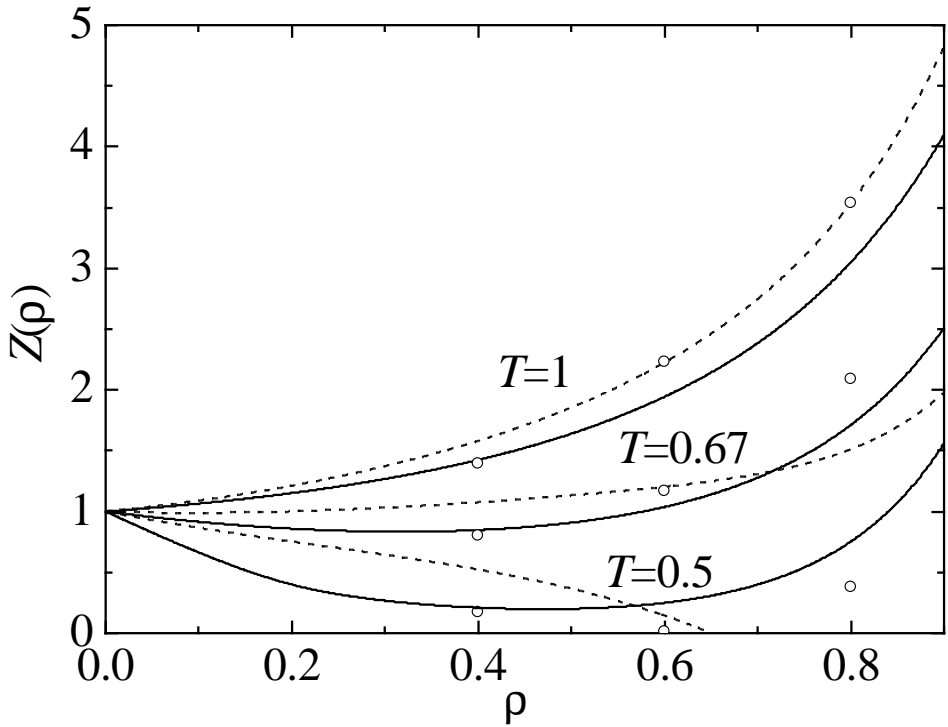}}
\caption{Density dependence of the compressibility factor for $\lambda=1.125$ 
and three temperatures. The circles are the results of computer 
simulations,\protect\cite{HSS80} the solid lines are the results from our 
model, and the dashed lines are the TL perturbation theory predictions.\cite{TL94b} 
\label{fig6}}
\end{center}
\end{figure}

\begin{figure}
\begin{center}\parbox{\textwidth}{\epsfxsize=0.7\hsize\epsfbox{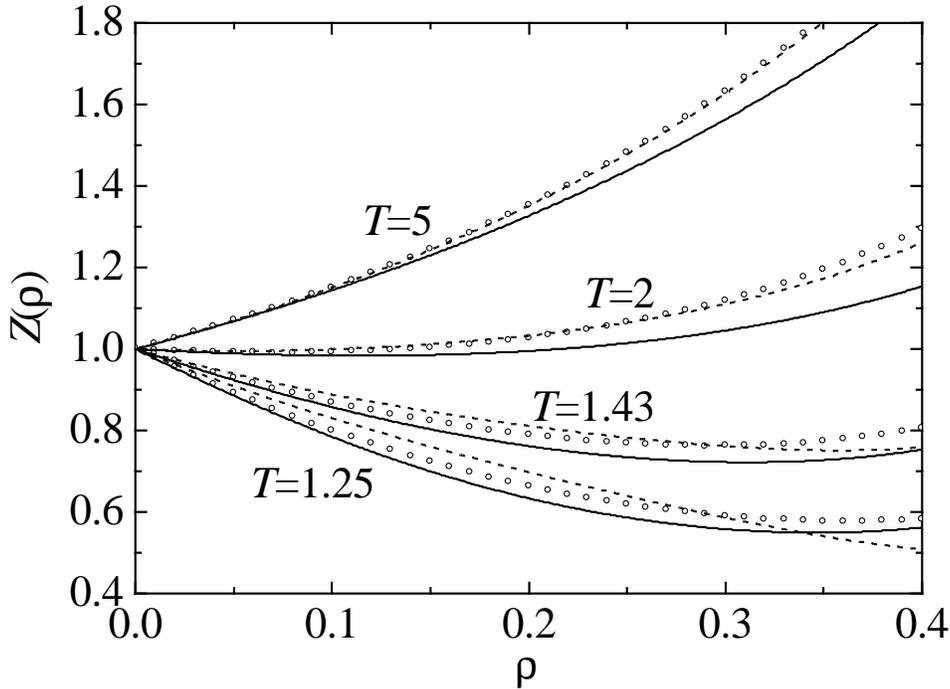}}
\caption{Density dependence of the compressibility factor for $\lambda=1.4$ 
and four temperatures. The circles correspond to an empirical formula fitted to
molecular dynamics results,\protect\cite{LLA90} the solid lines are the results from the 
present 
model, and the dashed lines are the TL perturbation theory predictions.\cite{TL94b} 
\label{fig7}}
\end{center}
\end{figure}

Now let us compare the general density dependence of the compressibility factor 
predicted by the model with available computer simulations. 
We start with the 
smallest value of $\lambda$ that, to our knowledge, has been analyzed in 
simulations, namely $\lambda=1.125$.\cite{HSS80} Figure \ref{fig6} shows 
$Z(\rho)$ for $\lambda=1.125$ and $T=0.5$, 0.67, and 1.
Strictly speaking, the curve representing the model at the lowest temperature
corresponds to $T=T_c\simeq 0.502$ rather than to $T=0.5$. This is because at
$T=0.5$ there exists a small density interval around $\rho_c\simeq 0.22$ [cf.\
Fig.\ \ref{fig5}] where the model gives complex values.
The next width we consider is $\lambda=1.4$, for which extensive molecular
dynamics simulations are available.\cite{LLA90} The results for $T=1.25$, 1.43,
2, and 5 are plotted in Fig.\ \ref{fig7}.  
We observe that, except at the highest temperatures ($T=1$ for $\lambda=1.125$, $T\geq 2$ for
$\lambda=1.4$), 
our model presents a better general agreement
with the simulation data than the TL perturbation theory. 
 
\section{Concluding remarks}
\label{sec_4}
The main objective of this paper has been to propose an analytical model that could be
useful to describe the structural and thermodynamic properties of systems, such
as colloidal dispersions,
composed of particles effectively interacting through a hard-core potential with
a short-range attraction. As the simplest interaction that captures both features
we have considered the square-well (SW) potential (\ref{a1}).
 On the one hand,
perturbation theory becomes unreliable when the range $\lambda-1$ of the
attraction is small, especially at low temperatures, i.e. when a certain degree of
``stickyness'' among the particles becomes important. On the other hand, the
widely used
sticky-hard-sphere (SHS) interaction model combines temperature and well width
in one parameter only, thus lacking the flexibility to accommodate additional 
changes in  width and/or temperature. Our approach intends to fill the gap
between these two theories. 

The model presented in this paper is based on the one proposed in Ref.\
\onlinecite{YS94}, where the functions $R(t)$ and $\overline{R}(t)$ defined by
Eqs.\ (\ref{b5}) and (\ref{c0}) were approximated by rational forms, Eq.\
(\ref{c1}). The
parameters in these functions are constrained to yield a finite value for the
isothermal compressibility by Eqs.\ (\ref{c5})--(\ref{c7bis}). 
This still leaves two free parameters, $A$ and $\tau$,
the latter being defined in Eq.\ (\ref{c10}), as unknown functions of the
packing fraction  $\eta$, the temperature parameter $x\equiv e^{1/T^*}-1$, and
the well width $\lambda$. Apart from their zero-density limits (\ref{c9})
 and (\ref{c12}), two extra conditions are needed to fix those parameters and
 close the construction of the model. Thus, depending on  the physical situation one is 
interested in and/or on the degree of simplicity one wants to keep in the model,
it is possible to choose different criteria to determine $A$ and $\tau$. For
instance, one could impose certain continuity conditions on the cavity function
$y(r)\equiv g(r)e^{\varphi(r)/k_BT}$ at the points where the potential is
singular;\cite{A00} alternatively, one could require thermodynamic consistency among the
virial, compressibility, and energy routes. Of course, other choices are
possible. In the original formulation of the
model, $A$ was assumed to be independent of density (hence $A=x$) and the second
condition was the continuity of $y(r)$ at $r=\lambda\sigma$, this giving rise to
a transcendent equation that needed to be solved numerically. On the other hand,
in this paper we have simply assumed that both $A$ and $\tau$ are independent of
density, Eq.\ (\ref{c13}). The assumption for $\tau$ is expected to be
especially adequate
for narrow potentials, since in the SHS limit the role of $\tau$ is played by the
parameter $\tau_{\text{SHS}}=[12x(\lambda-1)]^{-1}$, which is indeed independent
of density. With these choices for $A$ and $\tau$ the problem remains fully
algebraic and all the parameters can be expressed in terms of the solution of a
quadratic equation, in analogy with what happens in Baxter's solution of the PY
equation for SHS.\cite{B68} In fact, the model includes such a solution as a
limit case. 
Given the scarcity of simulation results for narrow wells, we
have been forced to carry out a comparison for cases with $\lambda\geq 1.125$.
In spite of that, the results show a very good general performance of the
structural and thermodynamic properties predicted by the model, correcting
the inadequacy of the perturbation theory predictions in the low-temperature
domain. It is interesting to remark that the model provides an explicit
expression of the critical temperature as a function of the well width, Eq.\
(\ref{d3}), which is  accurate even for rather wide wells.

\begin{figure}
\begin{center}\parbox{\textwidth}{\epsfxsize=0.7\hsize\epsfbox{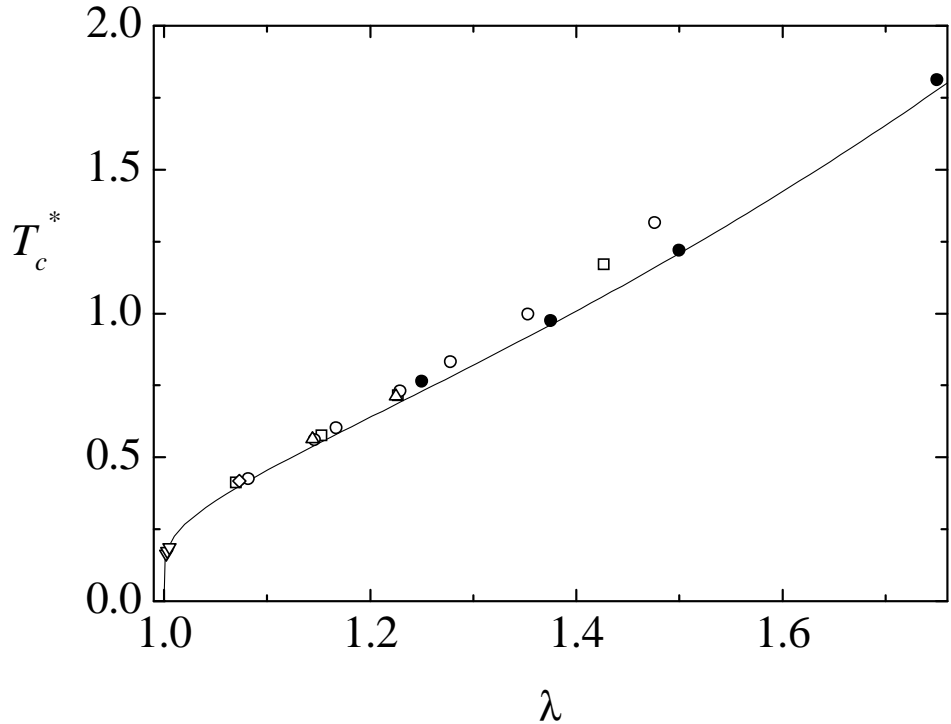}}
\caption{Dependence of the reduced critical temperature on the effective well
width, as obtained from computer simulations for several interaction
potentials: square-well (solid circles),\protect\cite{VMRJM92} hard-core Yukawa
(open squares),\protect\cite{NF00,LA94,HF94} generalized Lennard-Jones (open
circles),\protect\cite{NF00,VLL99} $\alpha$-Lennard-Jones (open diamond), 
\protect\cite{NF00,MA97}
an effective colloid-colloid interaction (up triangles), \protect\cite{NF00,MF94} and
an effective interaction for non-additive mixtures of asymmetric hard spheres
(down triangles).\protect\cite{NF00,D98,DRE99b} The solid line is the prediction of our
model, Eqs.\ (\protect\ref{d6}) and (\protect\ref{d3}).
\label{fig8}}
\end{center}
\end{figure}

The results of this paper could also be useful in connection with 
a recently proposed extension of the law of
corresponding states for systems, such as colloidal suspensions, that have
widely different range of attractive interactions.\cite{NF00} Given an interaction
potential $\varphi(r)=\varphi_{\text{rep}}(r)+\varphi_{\text{att}}(r)$, where
$\varphi_{\text{rep}}$ and $\varphi_{\text{att}}$ are the repulsive (not
necessarily hard-core)  and attractive parts, respectively, one can define a 
(temperature-dependent) effective hard-core diameter
\beq
\sigma=\int_0^\infty dr \,\left[1-e^{-\varphi_{\text{rep}}(r)/k_BT}\right],
\label{e1}
\eeq
an effective well depth
\beq
\epsilon=-\varphi(r_0),\quad \left.\frac{d\varphi(r)}{dr}\right|_{r=r_0}=0,
\label{e2}
\eeq
and a (temperature-dependent) effective well width
\beq
\lambda=\left[1+\left(B_2^*-1\right)\left(1-e^{\epsilon/k_BT}\right)^{-1}\right]^{1/3},
\label{e3}
\eeq
where
\beq
B_2^*=\frac{3}{\sigma^3}\int_0^\infty dr\, r^2\left[1-e^{-\varphi(r)/k_BT}\right]
\label{e4}
\eeq
is the reduced second virial coefficient. Then, according to the  extended law of corresponding
states,\cite{NF00} the compressibility factor for a wide range of colloidal
materials is a function of only the reduced temperature $T^*=k_BT/\epsilon$, the
reduced density $\rho^*=\rho\sigma^3$, and the range parameter $\lambda$, i.e.
$Z=F(T^*,\rho^*,\lambda)$, where the function $F$ is hardly sensitive to the
details of the potential. As a consequence, an accurate prescription for the
function $F$ based on the SW interaction for variable width can be used to
determine the thermodynamic properties of a wide class of colloidal suspensions.
As an illustrative example, Fig.\ \ref{fig8} shows the dependence of the
reduced critical temperature $T_c^*=k_BT_c/\epsilon$ on the effective width
$\lambda$ for several interaction potentials.\cite{NF00} The prediction of our
model, Eqs.\ (\protect\ref{d6}) and (\protect\ref{d3}), is also plotted.
The extended law of corresponding states work very well up to 
$\lambda\lesssim 1.3$. For larger interaction ranges the values of $T_c^*$ 
for the generalized
Lennard-Jones and the hard-core Yukawa potentials tend to lie slightly above
those corresponding to the SW interaction.

In the near future, we plan to extend the model presented in this paper in two
directions. First, the case of a mixture of particles interacting through SW
potentials with different values of $\sigma$, $\epsilon$, and $\lambda$ will
be analyzed. This generalization must be such that one recovers the cases of a
mixture of hard spheres, \cite{YSH98} a mixture of sticky hard spheres,
\cite{SYH98} and a monodisperse SW system in the appropriate limits. 
As a second extension, we will study the case of an interaction model made of a hard core
plus a square shoulder plus a square well, for which, in addition to the
conventional gas-liquid phase transition, a liquid-liquid  transition in the
supercooled phase appears.\cite{FMSBS00,MP00,FMSBS01}

\acknowledgments
 Partial support from the Ministerio de
Ciencia y Tecnolog\'{\i}a  (Spain) through grant No.\ BFM2001-0718
 and from the Junta de Extremadura (Fondo Social Europeo) through
grant No.\ IPR99C031 is gratefully acknowledged. A.S. is also grateful to the 
DGES (Spain) for 
a sabbatical grant No.\ PR2000-0117.
\appendix
\section{Some explicit expressions}
\label{appA}
\subsection{Expression for $S(0)$}
In this Appendix we list some of the expressions that are derived from the 
model. The long-wavelength value of the structure factor is obtained by using 
the model (\ref{c3}) in the general expression (\ref{b4}). The result is
\beqa
S(0)&=&\frac{1}{5(1+2\e)^2}\left\{5-20\e\left[1+(2+\l)A'-3(1+\l)L_2\right
]
\right.\nonumber\\
&&+2\e^2\left[15- 6(14-\l+19\l^2-\l^3-\l^4)L_2+120(1+\l+\l^2)L_2^2
\right.\nonumber\\
&&
+\left(50+16\l+27\l^2-2\l^3-\l^4\right)A'
-30(5+4\l+3\l^2)A'L_2\nonumber\\
&&\left.
+15(3+2\l+\l^2){A'}^2\right]
\nonumber\\
&&
-2\e^3\left[10+3(7-53 \l-13\l^2+7\l^3-8\l^4)L_2
\right.\nonumber\\
&&-60(1+\l)(1-4\l+\l^2)(1+\l+\l^2)L_2^2
\nonumber\\
&&
+\left(19+59\l+9\l^2-\l^3+4\l^4\right){A'}
\nonumber\\
&&
+3(11-67\l-114\l^2-66\l^3-13\l^4+9\l^5){A'}
L_2
\nonumber\\
&&
\left.
+3(3+2\l+\l^2)(1+7\l+3\l^2-\l^3){A'}^2\right]
\nonumber\\
&&
+\e^4\left[5-12(1+\l+\l^2+11\l^3-4\l^4)L_2
+240\l^3(1+\l+\l^2)L_2^2
\right.\nonumber\\
&&
+2\left(7+8\l+9\l^2+10\l^3-4\l^4\right){A'}-12(1+3\l+6\l^2+24\l^3+17\l^4+9\l^5)
{A'}L_2
\nonumber\\
&&
\left.\left.
+3(3+2\l+\l^2)(1+2\l+3\l^2+4\l^3){A'}^2\right]\right\},
\label{A1}
\eeqa
where we have made use of Eqs.\ 
(\ref{c5})--(\ref{c7bis}).

\subsection{Expressions for the coefficients in Eq.\ (\ref{c11})}
Equation (\ref{c10}) reduces to a quadratic equation for $L_2$. 
Its physical solution is given by Eq.\ (\ref{c11}), where
\beq
\alpha_1=2A'(2+\l)+\frac{1}{6}\left(1+4\l+\l^2-3A'\right)\tto,
\label{A2}
\eeq
\beq
\alpha_2=3+A'(7+2\l-3\l^2)-\frac{1}{12}\left[7+\l+16\l^2+A'
(23+15\l+15\l^2+7\l^3)\right]\tto,
\label{A3}
\eeq
\beq
\alpha_3=-2-2A'(1+2\l+3\l^2)+\frac{1}{6}\left[7+\l-2\l^2+A'
(7+15\l+21\l^2+11\l^3+6\l^4)\right]\tto
\label{A4}
\eeq
\beq
\beta_1=-4-4A'(2+\l)+\frac{1}{3}(5+2\l-\l^2+3A')\tto-\frac{1}{3}\kk2,
\label{A5}
\eeq
\beqa
\beta_2&=& 6+6A'(3+2\l+\l^2)+4{A'}^2(2+\l)^2
\nonumber\\
&&
-\frac{1}{6}\left[9(3+\l)+A' (59+51\l+3\l^2-5\l^3)+12{A'}^2 (2+\l)\right]\tto
\nonumber\\
&&
+\frac{1}{36}\left[31+8\l+18\l^2-4\l^3+\l^4+12A'(5+\l)+9{A'}^2\right]\kk2,
\label{A6}
\eeqa
\beqa
\beta_3&=&-4-12A'(1+\l+\l^2)-4{A'}^2(2+\l)(1+2\l+3\l^2)
\nonumber\\
&&
+\frac{1}{3}\left[3(4+\l+\l^2)+A'(29+36\l+30\l^2+7\l^3-3\l^4)
\right.
\nonumber\\
&&
\left.+{A'}^2 (17+43\l+30\l^2+\l^3-\l^4)\right]\tto
\nonumber\\
&&
-\frac{1}{36}\left[29+13\l+27\l^2+7\l^3-4\l^4+A' 
(68+80\l+80\l^2+16\l^3+7\l^4+\l^5)
\right.
\nonumber\\
&&
\left.+3{A'}^2(13+13\l+\l^2-3\l^3)\right]\kk2,
\label{A7}
\eeqa
\beqa
\beta_4&=&\left[1+A'(1+2\l+3\l^2)\right]^2
-\frac{1}{6}\left[7+\l+4\l^2
\right.
\nonumber\\
&&
\left.+2A'(7+15\l+15\l^2+5\l^3+6\l^4)+{A'}^2(1+2\l+3\l^2) 
(7+15\l+3\l^2-\l^3)\right]\tto
\nonumber\\
&&
+\frac{1}{144}\left[49+38\l+9\l^2+8\l^3+16\l^4+2A'(49+88\l+112\l^2+80\l^3 
+35\l^4-4\l^5)
\right.
\nonumber\\
&&
\left.+{A'}^2(49+138\l+219\l^2+124\l^3+27\l^4+18\l^5+\l^6)\right] \kk2,
\label{A8}
\eeqa
\beq
\gamma_0=1+\l-\frac{1}{6}\tto,
\label{A9}
\eeq
\beq
\label{A10}
\gamma_1=2(1+\l-\l^2)-\frac{1}{6}(3+\l^3)\tto,
\eeq
\beq
\gamma_2=\l\left[-4\l+\frac{1}{3}(2+3\l+\l^2+\l^3)\tto\right].
\label{A11}
\eeq
The other root is incompatible with (\ref{c9}) and then must be discarded.
\section{Low-density behavior of the model}
\label{appAB}
In the low-density limit, Eqs.\ (\ref{c5})-(\ref{c7bis}) become, respectively,
\begin{equation}
\label{a_1}
L_1=1-\frac{3}{2}\eta[1-x(\lambda^4-1)]+{\cal O}(\eta^2),
\end{equation}
\begin{equation}
\label{a_2}
S_1={\cal O}(\eta),
\end{equation}
\begin{equation}
\label{a_3}
S_2=-\frac{1}{2}[1-x(\lambda^2-1)]+{\cal O}(\eta),
\end{equation}
\begin{equation}
\label{a_4}
S_3=-\frac{1}{12\eta}+\frac{1}{3}[1-x(\lambda^3-1)]+{\cal O}(\eta).
\end{equation}
To first order in 
density, Eq.\ (\ref{c11}) gives
\beq
L_2=x\l(\l-1) +L_2^{(1)}\e+{\cal O}(\e^2),
\label{AB1}
\eeq
where
\beq
L_2^{(1)}=-\frac{3}{2}x\l(\l-1)\left\{1+x(\l-1)^2\left[1-2\l-\l^2-4x\l(1+\l)
\right]\right\}.
\label{AB2}
\eeq
Substitution into Eq.\ (\ref{c3}) yields, after some algebra,
\begin{eqnarray}
\label{a_7}
F(t)&=&F_{\text{exact}}(t)+\left\{\frac{2C_2}{t^4}+\frac{C_1}{t^3}+
\frac{C_0}{t^2} -\left[\frac{2C_2}{t^4}+\frac{C_1+2C_2(\lambda-1)}{t^3}
\right.\right.\nonumber\\
& &
\left.\left.
+\frac{C_0+C_1(\lambda-1)+C_2(\lambda-1)^2}{t^2}\right]e^{-(\lambda-1)t}\right\}
\eta+{\cal O}(\eta^2),
\end{eqnarray}
where we have called
\begin{equation}
\label{a_8}
C_2\equiv -3x(1+x)(\lambda^2-1),
\end{equation}
\begin{equation}
\label{a_9}
C_1\equiv 2x(1+x)(\lambda-1)^2(1+2\lambda),
\end{equation}
\beq
\label{a_10}
C_0\equiv \frac{L^{(1)}_2}{\lambda-1}-\frac{1}{2}x(2-8\l^3+3\lambda^4)
-2x^2(\lambda-1)^2(2+4\l+3\lambda^2)-3x^3(\lambda^2-1)^2.
\eeq
The expression for the exact radial distribution function,  
$g_{\text{exact}}(r)$, to first order in density was derived by Barker and 
Henderson for the case $\l<2$.\cite{BH67} The corresponding expression for 
$F_{\text{exact}}(t)$ can be found in Ref.\ \onlinecite{YS94}.
{}From Eq.\ (\ref{a_7}) one easily gets
\begin{equation}
\label{a_11}
g(r)-g_{\text{exact}}(r)=\frac{1}{r}[C_2(r-1)^2+C_1(r-1)+C_0]
[\Theta(r-1)-\Theta(r-\lambda)]\e+{\cal O}(\e^2).
\end{equation}
Thus, the difference (to first order in density) is  non-zero in the interval 
$1>r>\lambda$ only. In particular,
\beq
g(1^+)-g_{\text{exact}}(1^+)=C_0\e+{\cal O}(\e^2),
\label{AB3}
\eeq
where \cite{BH67}
\beqa 
g_{\text{exact}}(1^+)&=&1+x+\left[\frac{5}{2}+\frac{x}{2}(15-16\l^3+6\l^4) 
-2x^2(\l-1)(4+4\l+\l^2-3\l^3)
\right.\nonumber\\
&&\left.
+3x^3(\l^2-1)^2\right]\e+{\cal O}(\e^2).
\label{AB4}
\eeqa
Note that the relative coefficient $C_0/(1+x)$ vanishes in the SHS limit.
As said in Sec.\ \ref{sec_2}, the model presented in this paper does not enforce 
the verification of Eq.\ (\ref{c8}). In fact, Eq.\ (\ref{a_11}) implies that
\beqa
{g(\l^-)-(1+x)g(\l^+)}&=&{\l^{-1}}\left[C_2(\l-1)^2+C_1(\l-1)+C_0\right]\e+{\cal 
O}(\e^2)
\nonumber\\
&=&\frac{x(\l-1)}{2\l}\left[\l(11-\l-\l^2)-x(\l-1)(\l+2)(3+10\l-3\l^2)
\right.
\nonumber\\
&&\left.
+ 6x^2(\l-1)^2(1+3\l+2\l^2)\right]\e+{\cal O}(\e^2).
\label{AB5}
\eeqa

Finally, from Eq.\ (\ref{A1}) or, equivalently, inserting Eq.\ (\ref{a_11}) into 
Eq.\ (\ref{b1}), we get
\beqa
S(0)-S_{\text{exact}}(0)&=&2x(\l-1)^2\left[7+23\l+30\l^2-4\l^3-2\l^4-x(\l-1) 
\right.
\nonumber\\
&&\left.
\times (25+84\l+78\l^2+2\l^3-9\l^4)
+18x^2(\l^2-1)^2(1+2\l)\right]\e^2+{\cal O}(\e^3),
\label{AB6}
\eeqa
where
\beqa
S_{\text{exact}}(0)&=&1-8\left[1-x(\l^3-1)\right]\e
+2\left[17-x(\l-1)(19+19\l+19\l^2+51\l^3-3\l^4-3\l^5)\right.
\nonumber\\
&&
\left.
-2x^2(\l-1)^2(8+16\l-3\l^2-38\l^3-19\l^4)+18x^3(\l^2-1)^3\right]\e^2
+{\cal O}(\e^3).
\label{AB7}
\eeqa

\section{The hard-sphere and sticky-hard-sphere limits}
\label{appB}
\subsection{Hard spheres}
\label{appB1}
The SW potential becomes equivalent to the HS potential if $\l=1$ at any 
non-zero temperature $T$ or if $T\to\infty$ at any width $\l$. Let us first 
consider the latter limit. Making $x=0$ in Eqs.\ ({\ref{A2})--({\ref{A11}), one 
gets
\beq
\alpha_1=0,\quad \alpha_2=3,\quad \alpha_3=-2,
\label{B1}
\eeq
\beq
\beta_1=-4,\quad \beta_2=6,\quad \beta_3=-4,\quad \beta_4=1,
\label{B2}
\eeq
\beq
\gamma_0=1+\l,\quad \gamma_1=2(1+\l-\l^2),\quad \gamma_3=-4\l^2.
\label{B3}
\eeq
As a consequence, Eq.\ (\ref{c11}) implies that $L_2=0$. Thus Eq.\ (\ref{c3}) 
becomes 
\begin{equation}
\label{B4}
F(t)=-\frac{1}{12\eta}\frac{1+L_1 t}{1+
S_1t+S_2 t^2+S_3t^3},
\end{equation}
with
\begin{equation}
\label{B5}
L_1=\frac{1+\frac{1}{2}\eta}{1+2\eta},
\end{equation}
\begin{equation}
\label{B6}
S_1=-\frac{3}{2}\frac{\eta}{1+2\eta}
,
\end{equation}
\beq
\label{B7}
S_2=-\frac{1}{2}\frac{1-\e}{1+2\eta}
\eeq
\beq
\label{B8}
S_3=-\frac{(1-\eta)^2}{12\e(1+2\eta)}.
\eeq
This is precisely the form adopted by $F(t)$ in the analytical solution of the PY 
equation for hard spheres.\cite{W63,T63,YS91,YHS96} The expression for $S(0)$, 
Eq.\ (\ref{A1}), simply reduces to
\beq
S(0)=\frac{(1-\eta)^4}{(1+2\e)^2}.
\label{B9}
\eeq

The case $\l=1$ is not considered in this Subsection, as it is a particular case 
of the SHS limit.
\subsection{Sticky hard spheres}
\label{appB2}
Let us now take the limit $x\to\infty$, $\l\to 1$, with $x(\l-1)=\text{finite}$ 
in 
the model proposed in this paper. In that limit the parameter $\tau$ is finite, 
cf. Eq.\ (\ref{c13}), while $A'=0$. 
Equations ({\ref{A2})--({\ref{A11}) become
\beq
\alpha_1=\tto,\quad \alpha_2=3-2\tto,\quad \alpha_3=-2+2\tto,
\label{B10}
\eeq
\beq
\begin{array}{c}
\beta_1=-4+2\tto-\frac{1}{3}\kk2,\quad \beta_2=6-6\tto+\frac{3}{2}\kk2,\\
\beta_3=-4+6\tto-2\kk2,\quad \beta_4=1-2\tto+\frac{5}{6}\kk2,
\end{array}
\label{B11}
\eeq
\beq
\gamma_0=2-\frac{1}{6}\tto,\quad \gamma_1=2-\frac{2}{3}\tto,
\quad \gamma_3=-4+\frac{7}{3}\tto.
\label{B12}
\eeq
Therefore, Eq.\ (\ref{c11}) reduces to
\beq
L_2=\frac{1-\e}{24\e}\frac{(1+2\e) 
\left[(1-\e)^2+2\e(1-\e)\tto-\case{1}{6}\e(2-5\e)\kk2\right]^{1/2} 
-(1-\e)(1+2\e-\e\tto)}{ (1-\e)(1+2\e)-\case{1}{12}(1+4\e-14\e^2)\tto}.
\label{B13}
\eeq
Taking the limit $\l\to 1$ in Eq.\ (\ref{c3}) we get 
\begin{equation}
\label{B14}
F(t)=-\frac{1}{12\eta}\frac{1+L_1 t+L_2 t^2}{1+
S_1t+S_2 t^2+S_3t^3},
\end{equation}
with
\begin{equation}
\label{B15}
L_1=\frac{1+\frac{1}{2}\eta}{1+2\eta}+\frac{6\e}{1+2\e}L_2,
\end{equation}
\begin{equation}
\label{B16}
S_1=-\frac{3}{2}\frac{\eta}{1+2\eta}+\frac{6\e}{1+2\e}L_2
,
\end{equation}
\beq
\label{B17}
S_2=-\frac{1}{2}\frac{1-\e}{1+2\eta}+\frac{1-4\e}{1+2\e}L_2
\eeq
\beq
\label{B18}
S_3=-\frac{(1-\eta)^2}{12\e(1+2\eta)}-\frac{1-\e}{1+2\e}L_2.
\eeq
This coincides with the analytical solution of the PY equation for sticky hard 
spheres.\cite{B68,YS93a,YS93b} {}From Eq.\ (\ref{A1}) we have
\beq
S(0)=\frac{(1-\eta)^2}{(1+2\e)^2}\left[1-\e+12\e L_2\right]^2.
\label{B19}
\eeq
Of course, the results for hard spheres are recovered in the high-temperature 
limit ($\tto\to 0$).


\end{document}